# Emerging Topics in Internet Technology: A Complex Networks Approach

Bisma S. Khan[1], Muaz A. Niazi[*,2]
COMSATS Institute of Information Technology, Islamabad, Pakistan

*Abstract*—Communication networks, in general, and internet technology, in particular, is a fast-evolving area of research. While it is important to keep track of emerging trends in this domain, it is such a fast-growing area that it can be very difficult to keep track of literature. The problem is compounded by the fast-growing number of citation databases. While other databases are gradually indexing a large set of reliable content, currently the Web of Science represents one of the most highly valued databases. Research indexed in this database is known to highlight key advancements in any domain. In this paper, we present a Complex Network-based analytical approach to analyze recent data from the Web of Science in communication networks. Taking bibliographic records from the recent period of *2014* to *2017*, we model and analyze complex scientometric networks. Using bibliometric coupling applied over complex citation data we present answers to co-citation patterns of documents, co-occurrence patterns of terms, as well as the most influential articles, among others, We also present key pivot points and intellectual turning points. Complex network analysis of the data demonstrates a considerably high level of interest in two key clusters labeled descriptively as "social networks" and "computer networks". In addition, key themes in highly cited literature were clearly identified as "communication networks," "social networks," and "complex networks".

## I. INTRODUCTION

Modern communication networks can rightfully be considered the harbinger of change. Wherever in society, there are computers and networks, changes can be observed to take place. Technological advancements allow for inventions and applications which could previously only have been considered as science fiction.

One key problem with communication systems and networks is the fast-changing literature. As new areas and terms keep emerging every other day, it can be very difficult to keep track of every gradual advancement in one's particular sub-domain. To make matters more complex, at times, some topics emerge in social media however they are not really considered important by related researchers. Whereas at other times, some areas might be emerging topics but practitioners remain unaware of their importance.

A key reason this is a complex problem is because citations and bibliographic coupling are essentially a Complex Adaptive System (CAS). Modeling the system as CAS allows for modeling the different aspects of the system. Typical techniques of modeling CAS involve using agents [1] or else complex networks [2].

While previous work has used the Web of Science and Citation networks to identify patterns in agent-based computing [3], modeling and simulation [4], and consumer electronics [5], to the best of our knowledge, the same techniques have not been applied to the domain of Internet domain.

In this paper, we applying Cognitive Agent-based Computing framework for modeling CAS [6] on data from the Web of Science using CiteSpace [7].

The key contribution of this article is the identification of important nodes including both highly cited as well as most central articles, citation bursts, article clusters, turning points, pivot points, key indexing terms among others.

The rest of the paper is organized as follows: Section II presents the methodology, Section III demonstrates results, and Section IV contains conclusions.

## II. METHODOLOGY

We have retrieved input data from Clarivate Analytics' Web of Science (WoS) on Mar *18, 2017*. An exhaustive topic search was performed using a query "TS = ((((("Communication") OR ("Telecommunication")) AND ("Network*")) OR ("Complex* Network*"))" in the timespan of *2014-2017*. We have selected all English-language documents. The bibliographic data includes the full record and cited references. The search was performed on all four indices of WoS, including SCI-EXPANDED, SSCI, A&HCI, ESCI and is refined by excluding some of the Biology, Psychology, and Chemistry related categories. The search resulted in *22,944* unique records with *37.042%* records from the year *2016*; *32.305%* records from *2015*; *27.144%* records from *2014*; and *3.509%* records from *2017*. We have used CiteSpace, a network analytic tool for network analysis.

[1]bis.sarfraz@gmail.com, *Corresponding author: [2]muaz.niazi@ieee.org

## III. RESULTS

Several visualizations were used to identify latest developments, emerging trends, and temporal patterns in the domain.

### A. Identification of Emerging Trends and Topics

The first analysis was to analyze Google trends and Carrot search FoamTree visualization. Google trends in Figure 1 presents search interest of the topics: communication networks (blue line), complex networks (red line), telecommunication networks (yellow line), computer networks (green line), and social networks (purple line), relative to the total volume of "social networks" remained most popular over a long time, but "Computer Networks" leads over it from Jul *19, 2016*, onward. Apparently "social networks" remained most popular over a long time, but "Computer Networks" leads over it from Jul *19, 2016*, onward.

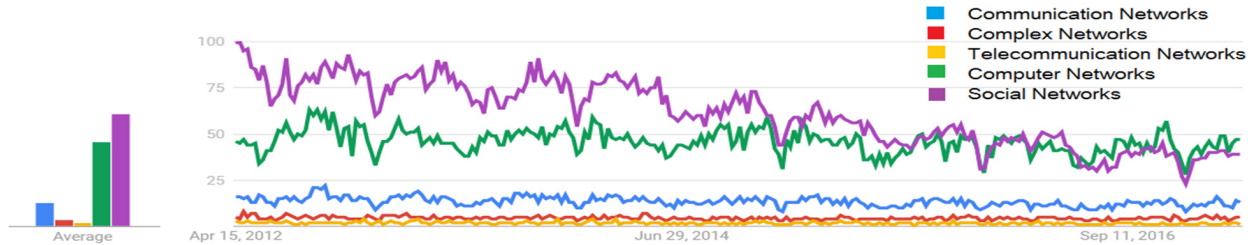

Figure 1. Google trends demonstrating search interest over past five years

Figure 2 depicts the major topics of the domain. Carrot Search FoamTree visualization as depicted in Figure 2 is generated based on *200* web search results clustered with Lingo3G, given the query: "((("Complex") OR ("Computer") OR ("Social") OR ("Communication") OR ("Telecommunication")) AND ("Network*"))". "Communication Networks" is the leading topic in the visualization. Other important clusters are "Social Networks" and "Complex Networks."

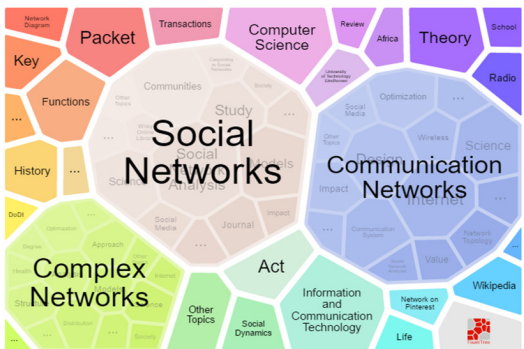

Figure 2. Carrot Search FoamTree visualization based on 200 web results

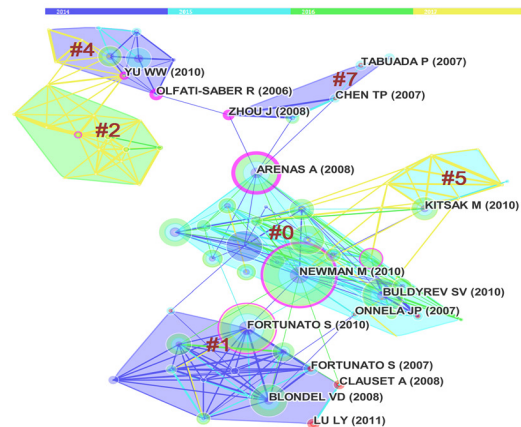

Figure 3. Largest connected cluster based on cited references

### B. Cited Reference Co-Citation Network Analysis

This section presents co-citation analysis of the cited reference network. The panoramic network contains *112* cited references and *320* co-citation links. We have selected the top *50* documents per time slice of one-year length for the time period of *2014* to *2017*. Each slice corresponds to a unique color. Link colors also represent a specific time slice. Figure 4 illustrates clusters in the Largest Connected Component (LCC) of the cited reference co-citation network. The LCC contains *78* nodes, which constitute *69%* of the network. The largest radius of Newman M (*2010*) represents the highest citation frequency. The thickness of purple trims around Arenas A (*2008*) represents the highest betweenness centrality of the article. The pink circles around the nodes are highlighting the articles with a centrality score $>= 0.1$.

The LCC as shown in Figure 3 contains six clusters (*#0, #1, #2, #4, #5*, and *#7*). The cluster labels are the noun phrases extracted from the title, index terms, or an abstract of the paper. They represent underlying theme of the cluster. The LCC has modularity score of *0.718* which indicates the relatively high network density. The silhouette of *0.5114* represents average homogeneity amongst members of the clusters. Most of the burst articles belong to cluster *#1* and others belong to Cluster *#0* and *#7*, which shows that these three clusters are the active areas of the domain.

The articles Olfati-Saber R (*2006*) and Zhou J (*2008*) with prominent purple trims are the turning points which connect cluster *#4* and cluster *#7*. The article Arenas A (*2008*) is the pivot point joining cluster *#0* and cluster *#7*. These nodes also serve as bridges in joining different clusters.

3As shown in Table II, the largest cluster (cluster #0) contains *25* nodes, which are *22.32%* of the entire network and *32.05%* of the LCC. The most influential article, Newman M (*2010*) and the most central article Arenas A (*2008*) belong to this cluster. It has relatively high mean silhouette score of *0.7*, which is the indicator of high homogeneity among members of the cluster.

The second largest cluster (cluster #1) contains *15* nodes, which are *19.23%* of the LCC and *13.39%* of the entire network. It is a highly homogeneous cluster with a silhouette score of *0.939*. Mean publication year of this cluster is *2008*. The most active article by Fortunato S (*2010*) also belongs to this cluster. Other articles Costa LD (*2007*), Lu LY (*2011*), and Clauset A (*2008*) with highest citation burst also belong to this group. It indicates that it is the most active area of the domain and is the focus of the interest of the research community.

Table III demonstrates top documents in terms of frequency, centrality, and burstness. The first column of Table III lists top five cited articles sorted in terms of frequency of citations.

TABLE I. CLUSTERS IN THE LARGEST CONNECTED COMPONENT OF CITED REFERENCES

| CID | Size | Sihouette | Mean Year | Label (TFIDF) | Label (LLR) | Label (MI) |
|---|---|---|---|---|---|---|
| 0 | 25 | 0.7 | 2010 | Normal Plasma Vasopressin \| Pulmonary Artery | Complex Network; Interdependent Network; Free Network; | Digital Footprint |
| 1 | 15 | 0.939 | 2008 | Complex Network \| Systems Biology | Community Detection; Community Structure; Complex Network; | Neighborhood Coreness |
| 2 | 14 | 0.868 | 2012 | Transmission Delay \| Event-Triggered Fault Detection Filter Design | Networked Control System; Complex Network; H Infinity Control; | Diffusion LMM |
| 4 | 10 | 0.848 | 2008 | Mobile Sensor Network \| Matrix Algebra | Multi Agent System; Multi-Agent System; Complex Network; | Information Consensus |
| 5 | 9 | 0.927 | 2011 | Structural Perturbation \| Diffusion Game | Influential Spreader; Influential Node; Spreading Probability; | Complex Network |
| 7 | 5 | 0.971 | 2007 | Transmission Delay \| Networked System | Networked Control System; Pinning Control; Pinning Synchronization; | Actuator Network |

The book by Mark J Newman (*2010*) in cluster #0, is the most influential node of the domain with a frequency of *542* citations. It has a half-life [8] of five years. Next, we have an article by Santo Fortunato (*2010*) in cluster #2, with *435* citations. It has a HL of five years and has *5199* citations on Google Scholar (GS). Then we have an article by Sergey V Buldyrev (*2010*) in cluster #0 with *321* citation frequency. It has a HL of five years and has *1762* citations on GS. Next is the article by Alex Arenas (*2008*) in cluster #0, with *319* citations. It has a HL of seven years and has *1920* citations on GS. Finally, we have an article by Vincent D Blondel (*2008*) in cluster #2 with *281* citations. It has a HL of seven years and has *5026* citations on GS.

The third column in Table III presents top five authors sorted in terms of centrality. The article by Alex Arenas (*2008*) is on top with a centrality score of *0.46*. Next is an article by Jin Zhou (*2008*) in cluster #7 with *0.42* centrality score. It has a HL of six years and has *473* citations on GS. Then, we have an article by Reza Olfati-Saber (*2006*) in cluster #3 with *0.41* centrality score. It has a HL of eight years and has *3127* citations on GS. Next, we have an article by Wenwu Yu (*2010*) in cluster #3 with a centrality score of *0.3*. It has a HL of four years and has *643* citations on GS. Subsequently, we have a book by Mark J Newman (*2010*) with a centrality score of *0.27*.

TABLE II. TOP ARTICLES SORTED IN TERMS OF FREQUENCY, CENTRALITY, AND BURSTNESS

| Article | Frequency | Article | Centrality | Article | Citation Burst | | | |
|---|---|---|---|---|---|---|---|---|
| | | | | | Strength | Begin | End | 2014 – 2017 |
| Newman M [9] | 542 | Arenas A [10] | 0.46 | Fortunato S [11] | 8.73 | 2014 | 2015 | |
| Fortunato S [12] | 435 | Zhou J [13] | 0.42 | Tabuada P [14] | 7.05 | 2014 | 2015 | |
| Buldyrev SV [8] | 321 | Olfati-saber R [15] | 0.41 | Boyd DM [16] | 6.63 | 2014 | 2015 | |
| Arenas A [10] | 319 | Yu WW [17] | 0.3 | Chen TP [18] | 6.55 | 2014 | 2015 | |
| Blondel VD [19] | 281 | Newman M [9] | 0.27 | Onnela JP [20] | 6.21 | 2014 | 2015 | |

The fifth column in Table III contains top five documents based on the burst of citations. The article by Fortunato S (*2007*) has strongest citation burst of *8.73* from *2014* to *2015*. Next, we have an article by Paulo Tabuada (*2007*) in cluster #7 with a citation burst of *7.05* from *2014* to *2015*. It has a HL of seven years and it has *1239* citations on GS. Then we have an article by Danah M Boyd (*2007*) in cluster #11, having citation burst of *6.63* from *2014* to *2015*. It has a HL of eight years and has *11794* citations on GS. Next is an article by Chen Tianping in cluster #7, having citation burst of *6.55* from *2014* to *2015*. It has a HL of eight years and has *643* citations on GS. Subsequently, we have an article by Jukka-Pekka Onnela (*2010*) in cluster #0, with a citation burst of *6.21* from *2014* to *2015*. It has a HL of seven years and has *1366* citations on GS.

*C. Term Co-Occurrence Network Analysis*

We carried a visualization of terms network to find the key indexing terms of the domain. The merged network of terms contains *89* nodes and *299* links as shown in Figure 4.

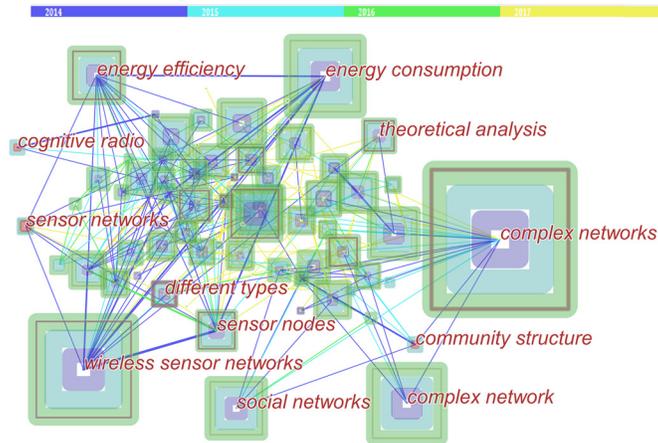

Figure 4. A visualization of the core terms retrieved from bibliographic records

We have selected top *50* nodes per one-year time slice in the timespan of *2014-2017*. The term "Complex networks" with thickest purple trim is the most central node. The pink trims around vertices is the indicator of the betweenness centrality score >= *0.1*. The largest area of the square around the term "complex networks" indicates that it is the landmark node. Red marks on the nodes represent burstness, which indicates a sudden change of the frequency of citations over time in a specific duration. Red squares on "cognitive radio" indicates considerable attention of the scientific community in a particular time frame. The concentric squares around vertices represent the citation history of the publications over time. The color of the citation square represents the citation in a specific time slice. The link color represents the time slice when the link was first created. A comprehensive analysis is given below in the tabular form.

The first column in Table I contains top five indexing terms sorted in terms of frequency. The term "complex networks/complex network" leads over other indexing terms of the domain with the highest citation frequency of *2154*. The third column in Table I lists top five indexing terms based on centrality score. The "complex networks" is the most central indexing term with a centrality score of *0.31*. The sixth column in Table I presents five key indexing terms in terms of citation burst in the timespan of *2014-2017*. The "cognitive radio" is the most active term of the domain with strongest citation burst of *14.4* which last from *2014* to *2015*.

TABLE III. TOP INDEXING TERMS BASED ON FREQUENCY, CENTRALITY, AND BURSTNESS.

| Terms | Frequency | Terms | Centrality | Terms | Citation Burst Begin | End | Strength | 2014 – 2017 |
|---|---|---|---|---|---|---|---|---|
| Complex Network(s) | 2154 | Complex Networks | 0.31 | Cognitive Radio | 2014 | 2015 | 14.4 | |
| Wireless Sensor Networks | 1012 | Different Types | 0.2 | Sensor Networks | 2014 | 2015 | 14.39 | |
| Energy Consumption | 740 | Energy Efficiency | 0.16 | Community Structure | 2014 | 2015 | 13.5 | |
| Social Networks | 557 | Sensor Nodes | 0.15 | Computational Complexity | 2015 | 2017 | 9.56 | |
| Energy Efficiency | 533 | Theoretical Analysis | 0.13 | Numerical Example | 2015 | 2017 | 6.53 | |

## IV. CONCLUSIONS

In this study, CiteSpace has been used to explore emerging trends and developments in internet technology. Toward that end, we carried a comprehensive visualization of bibliographic literature retrieved from Clarivate Analytics' Web of Science in the time frame of *2014 – 2017*. We have identified bursting, landmark, and influential terms and articles, cited reference co-citation network, terms collaboration network, turning points, pivot nodes, and trendy topics.

Our analysis revealed that the indexing term "cognitive radio" has received remarkable attention from the research community from *2014* to *2015*. "Complex network" is the highly cited and most central key indexing term. The terms "computational complexity" and "numerical examples" has longest citation burst lasting from *2015* to date. The largest connected cluster of the cited references contains *25* members which are *22.32%* of the entire network. The article Fortunato S (*2007*) has received a surge of citations from *2014* to *2015*. It has received *1691* citations on Google Scholar. H-index of Santo Fortunato is *37* since *2012*.

Besides conclusions, we believe that our analysis will provide a broader picture of the domain to the research community.